\documentclass[fleqn,usenatbib]{mnras}
\usepackage{newtxtext,newtxmath}
\usepackage{textgreek}

\usepackage[T1]{fontenc}
\usepackage{ae,aecompl}

\usepackage{graphicx}   
\usepackage{amsmath}    
\usepackage{amssymb}    

\usepackage[para, online, flushleft]{threeparttable}

\title[Astrometric requirements for strong lensing time-delay cosmography]{Astrometric requirements for strong lensing time-delay cosmography}

\author[S. Birrer]{
Simon Birrer$^{1}$\thanks{E-mail: sibirrer@astro.ucla.edu}
and Tommaso Treu$^{1}$
\\
$^{1}$Department of Physics and Astronomy, University of California, Los Angeles, CA 90095, USA\\
}

\date{Accepted XXX. Received YYY; in original form ZZZ}

\pubyear{2019}

\begin{document}
\label{firstpage}
\pagerange{\pageref{firstpage}--\pageref{lastpage}}
\maketitle

\begin{abstract}
The time delay between the arrival of photons of multiple images of time variable sources can be used to constrain absolute distances in the Universe (Refsdal 1964), and in turn obtain a direct estimate of the Hubble constant and other cosmological parameters. To convert the time delay into distances, it is well known that the gravitational potential of the main deflector and the contribution of the matter along the line-of-sight need to be known to a sufficient level of precision.
In this paper, we discuss a new astrometric requirement that is becoming important as time-delay cosmography improves in precision and accuracy with larger samples, and better data and modelling techniques. We derive an analytic expression for the propagation of astrometric uncertainties on the multiple image positions into the inference of the Hubble constant and derive requirements depending on image separation and relative time delay. We note that this requirement applies equally to the image position measurements and to the accuracy of the model in reproducing them. To illustrate the requirement, we discuss some example lensing configurations and highlight that, especially for time delays of order 10 days or shorter, the relative astrometric requirement is of order milli-arcseconds, setting a tight requirement on both measurements and models. With current optical infrared technology, astrometric uncertainties may be the dominant limitation for strong lensing cosmography in the small image-separation regime when high-precision time-delays become accessible.

\end{abstract}




\section{Introduction}
The time delay between the arrival of photons of multiple images of gravitationally lensed sources provides a physical anchor of the scales in the universe. This method, known as time-delay cosmography or time-delay strong lensing  \citep{Refsdal:1964pi}, provides a one-step measurement of the Hubble constant and other cosmological parameters. Time-delay cosmography is independent of other cosmological probes \citep{Treu:2016}, such as the Cosmic Microwave Background \citep[CMB;][]{Planck2018_cosmo_param} or the local distance ladder \citep{Riess:2019, Freedman:2019}.

In the past decades, measurements of relative time delays have been achieved with lensed variable active galactic nuclei (AGNs), requiring multi-year monitoring campaigns with per-cent-level photometry \citep{Fassnacht:2002, COSMOGRAIL_1:2005, Kochanek:2006, Tewes:2013xr, Liao:2015, Bonvin:2016, tak2017bayesian} or high-cadence monitoring with millimag photometry \citep{Courbin:2018, Bonvin:2018a} with uncertainties of $\sim$ 1 day. These galaxy-scale AGN lenses have image separations of 1\arcsec - 3\arcsec with relative delays of about 10-100 days from the first and last image appearing, lensed by a massive early-type galaxy. More recently, lensed supernovae, as originally suggested by \citet{Refsdal:1964pi}, have been discovered \citep{Kelly:2015,Treu:2016Refsdal, Goobar:2017} and may become an important source of time-delay cosmography in their own right \citep{Goldstein:2017, Grillo:2018}. Other classes of multiply imaged time-variable sources may be observed and analyzed to perform cosmographic measurements, such as repeating Fast Radio Bursts \citep{Li:2018_FRB} or gravitational waves \citep{Sereno:2011_GW_cosmography}. In this paper we choose lensed AGNs as an illustration, but we note that the astrometric requirement equally applies to transient sources, with the caveat that opportunities to gather high precision astrometric measurements will be limited to the visibility window.

In addition to a measurement of the difference in arrival time, the other ingredients for time delay cosmography are: 1) a precise model of the gravitational potential across the images to estimate the relative Fermat potential; 2) the line-of-sight weak lensing effect that can alter the cosmographic distances. Fermat potential and line-of-sight effect need to be inferred with independent data, such as high-resolution \textit{Hubble Space Telescope} (\textit{HST}) imaging of the lensing arc, imaging and spectroscopy of the environment of the lens and kinematics of the lensing galaxy \citep[see e.g.][]{Suyu:2017_h0licow}.

Modern state-of-the-art cosmographic analysis has been performed to date on 6 galaxy-scale AGN lenses \citep{Suyu:2010rc, Suyu:2013ni, Suyu:2014aq, Birrer:2016zy, Wong:2017, Bonvin:2017, Birrer:2019_1206, Rusu:2019, Chen:2019} yielding uncertainties of order 5-8 per cent per system, resulting in a joint measurement of the Hubble constant with 2.4 per cent precision \citep{Wong:2019}.

Time-delay cosmography has the potential to measure the Hubble constant to one per cent precision in the near future with a sample of about 40 lenses \citep{Treu:2016, Jee:2016, Shajib:2018}, a forecast based on the demonstrated precision in current analyses of lensed AGNs.

In this paper, we discuss an additional requirement on the astrometric precision of the relative positions of the images of the time-variable source. The relative astrometric uncertainties may lead to substantial cosmological uncertainties and possibly biases if neglected, and has not been sufficiently discussed in the literature. 

We derive an analytic expression for the propagation of astrometric uncertainties into the inference of the Hubble constant and derive requirements depending on image separation and relative time delay.
We test this expression with numerical examples where we displace the image positions. We derive requirements for different types of lenses and sources such that the astrometric uncertainty is sub-dominant with respect to other uncertainties inherent in the cosmographic analysis.

We discuss the different regimes in which time-delay cosmography is applied and elaborate on the expected impact on current and future analyses. Typically, not to dominate the cosmological error budget, the relative astrometry of the variable multiple images needs to be known with $\lesssim 10$ mas uncertainty. For short time delays ($\lesssim 10$ days), the astrometric precision needs to be $\lesssim 1$ mas.
We note that the astrometric requirements discussed in this work go beyond statements about the data itself and are equally required by the modelling aspects, e.g. in the crowded regime when multiple light components are super-imposed, and to the accuracy of the model in reproducing the observed image positions.

The paper is structured as follow: In Section \ref{sec:theory}, we review the theory of gravitational lensing and its application to cosmography with time-variable sources. In Section \ref{sec:astrometric_error_propagation}, we propagate the astrometric error into the cosmographic analysis and derive an analytic expression applicable terms of relative time delays and image separations. In Section \ref{sec:specific_requirements}, we provide examples for different and illustrative lensing systems and derive the astrometric requirements for these cases. In Section \ref{sec:conclusion}, we elaborate on the implications for current and future studies of time variable lensing systems.

All numerical computations are performed with \textsc{lenstronomy}\footnote{\url{https://github.com/sibirrer/lenstronomy}} \citep{Birrer2015_basis_set, Birrer_lenstronomy} version \textsc{0.7.0}.

\section{Time-delay cosmography} \label{sec:theory}
In this section, we provide a basic review of the lensing theory with a focus on time-delay cosmography. For conciseness, we do not discuss lensing degeneracies or the effect of multiple lens-planes in this paper and refer the reader to the current literature in this regard \citep[e.g.][]{Schneider:2019}.

\subsection{Lensing formalism and time delays}
The lens equation, which describes the mapping from the source plane $\vec{\beta}$ to the image plane $\vec{\theta}$ is given by
\begin{equation} \label{eqn:lens_equation}
  \vec{\beta} = \vec{\theta} - \vec{\alpha}(\vec{\theta}),
\end{equation}
where $\vec{\alpha}$ is the angular shift on the sky between the image position it had in absence of the deflector and the actual observed position. The vector field $\vec{\alpha}$ can be derived from a scalar potential $\psi$, known as the lensing potential, such that
\begin{equation}
    \vec{\alpha}(\vec{\theta}) = \vec{\nabla} \psi(\vec{\theta}).
\end{equation}
The lensing potential is related to the surface mass density by the two-dimensional Poisson equation.

The excess time delay of an image at position $\vec{\theta}$ relative to the unperturbed path is
\begin{equation}\label{eqn:time-delay}
 t(\vec{\theta}, \vec{\beta}) = \frac{(1 + z_{\text{d}})}{c} \frac{D_{\text{d}}D_{\text{s}}}{D_{\text{ds}}} \left[ \frac{(\vec{\theta} - \vec{\beta})^2}{2} - \psi(\vec{\theta}) \right],
\end{equation}
where $z_{\rm d}$ is the redshift of the deflector, $c$ the speed of light, $\psi$ the lensing potential and $D_{\rm d}$, $D_{\rm s}$ and $D_{\rm ds}$ the angular diameter distances from the observer to the deflector, from the observer to the source and from the deflector to the source, respectively.
The angular term in brackets in the equation above, a combination of the geometrical and the gravitational delays, is called the Fermat potential $\phi$ 
\begin{equation}
  \phi(\vec{\theta}, \vec{\beta}) \equiv \left[ \frac{(\vec{\theta} - \vec{\beta})^2}{2} - \psi(\vec{\theta}) \right].
\end{equation}

The relative time delay between two images A and B, $\Delta t_{\rm AB}$, is
\begin{equation} \label{eqn:td_formula}
    \Delta t_{\rm AB} = \frac{D_{\Delta t}}{c} \Delta \phi_{\rm AB},
\end{equation}
where
\begin{equation} \label{eqn:time_delay_distance}
 D_{\Delta t} \equiv (1+z_{\text{d}})\frac{D_{\text{d}}D_{\text{s}}}{D_{\text{ds}}}
\end{equation}
is the so-called time-delay distance
and 
\begin{equation}
\Delta \phi_{\rm AB} \equiv \phi(\vec{\theta}_{\rm A}, \vec{\beta}) - \phi(\vec{\theta}_{\rm B}, \vec{\beta}).
\end{equation}
is the relative Fermat potential between two images.

\subsection{Cosmography and the Hubble constant}

A measurement of the relative time delay, $\Delta t_{\rm AB}$, and the relative Fermat potential, $\Delta \phi_{\rm AB}$ allows one to infer the time-delay distance, $D_{\Delta t}$ (Equations \ref{eqn:td_formula} and \ref{eqn:time_delay_distance}). Conceptually, whereas all the other lensing observables are angles, and thus do not contain distance information, the time delay multiplied by the speed of light is an absolute measurement of length and can thus measure distances.

In practice, the time delay is currently measured with ground based monitoring campaigns and the relative Fermat potential is estimated from the distortion observed in high-resolution images and the velocity dispersion measurements of the lensing galaxy (see e.g. \cite{Suyu:2017_h0licow} and references therein).

The time-delay distance is the primary cosmographic measurement and anchors the absolute scale of the universe for the specific redshift configuration of the lensing system. Like for any absolute distance measurement in the Hubble flow, the Hubble Constant (H$_0$) is inversely proportional to the distance ($D_{\Delta t}$ in the case of lensing). 

Thus, the relative error contribution in $D_{\Delta t}$, and thus H$_0$, from a time delay measurement $\Delta t \pm \delta \Delta t$ is directly linked to the relative error in the time delay as

\begin{equation} \label{eqn:hubble_dt_error}
  \frac{\delta {\rm H}_0}{{\rm H}_0} = \frac{\delta \Delta t}{\Delta t}
\end{equation}
and the error on the inference of the relative Fermat potential $\Delta \phi_{\rm AB} \pm \delta \Delta \phi_{\rm AB}$
\begin{equation} \label{eqn:hubble_fermat_error}
  \frac{\delta {\rm H}_0}{{\rm H}_0} = 
\frac{\delta \Delta \phi_{\rm AB}}{\Delta \phi_{\rm AB}}.
\end{equation}

\section{Astromeric error propagation} \label{sec:astrometric_error_propagation} 
Astrometric errors in the determination of an image position $\vec{\theta}_{\rm A}$ affect the estimate of the relative Fermat potential, $\Delta \phi_{\rm AB}$ (Eqn \ref{eqn:time-delay}), regardless of the precision of the lensing potential, $\psi$. Therefore, astrometric errors can be a limiting factor to the achieved cosmographic precision.

In this section, we perform the first order error propagation (Section \ref{sec:astrometry_first_order}) and then a full non-linear propagation (Section \ref{sec:astrometry_non_linear}), showing that the first order term is the dominant one.
We then derive a simplified form of the error propagation in terms of an observed time delay and image separation that can be used to derive the specific requirements on the relative astrometry for any given time-delay lensing system (Section \ref{sec:astrometry_requirement}).

\subsection{Linear error propagation} \label{sec:astrometry_first_order}
The first order correction to the relative Fermat potential difference, $\Delta \phi_{\rm AB}$, by a displacement $\delta \vec{\theta}_{\rm A}$ is given by

\begin{multline}\label{eqn:first_order}
\delta \Delta \phi_{\rm AB}(\delta \vec{\theta}_{\rm A}) \approx 
\frac{d \Delta \phi_{\rm AB}}{d \vec{\theta}_{\rm A}} \cdot \delta \vec{\theta}_{\rm A}
 = \left[\frac{\partial}{\partial \vec{\theta}_{\rm A}} + \frac{\partial \vec{\beta}}{\partial \vec{\theta}_{\rm A}} \frac{\partial}{\partial \vec{\beta}}\right] \Delta \phi_{\rm AB} \cdot \delta \vec{\theta}_{\rm A}\\
= \left[ \vec{\theta}_{\rm A} - \vec{\beta} -\vec{\alpha}(\vec{\theta}_{\rm A}) + \frac{\partial \vec{\beta}}{\partial \vec{\theta}_{\rm A}}  \left(\vec{\theta}_{\rm B} - \vec{\theta}_{\rm A} \right) \right] \cdot
\delta \vec{\theta}_{\rm A}\\
 = \left(\vec{\theta}_{\rm B} - \vec{\theta}_{\rm A} \right) \frac{\partial \vec{\beta}}{\partial \vec{\theta}_{\rm A}}
\delta \vec{\theta}_{\rm A},
\end{multline}
where we have applied the lens equation (Eqn \ref{eqn:lens_equation}) in the last line of the equation above. We recover the fact that lensed images appear at extrema of the arrival time surface (Fermat potential) and thus the partial differentials of the Fermat potential at the image positions vanish to first order. The remaining first order term is the partial derivative of the source position with respect to a shift in the image position. This calculation effectively updates the source position based on the displacement of the position of image A only. In reality, each individual image provides constraints on the source positions.

In a more general way, we can express Equation~\ref{eqn:first_order} as

\begin{equation} \label{eqn:error_vector}
\delta \Delta \phi_{\rm AB}(\delta \vec{\theta}) \approx \left(\vec{\theta}_{\rm B} - \vec{\theta}_{\rm A} \right) \cdot \delta \vec{\beta},
\end{equation}
where we introduced the notation $\delta \vec{\beta}$ for the propagated astrometrical error on the source plane from the combined information available for each individual image. We note that this expression is a scalar product and has a directional dependence. Displacements of the source along the direction between the two images have the most significant impact on the relative Fermat potential. \footnote{\cite{Wagner:IV2018} derived Equation \ref{eqn:error_vector} as the general condition to connect the transformation of the deflection potential with a shift of the source position for given image positions.}

The variance in the relative Fermat potential, $\sigma^2_{\Delta \phi_{\rm AB}}$, can be expressed in terms of an error covariance matrix in the source plane position, $\boldsymbol{\Sigma}_{\beta} \equiv \sigma(\beta_i, \beta_j)$, as

\begin{equation} \label{eqn:linear_fermat_errors}
\sigma^2_{\Delta \phi_{\rm AB}} \approx \left(\vec{\theta}_{\rm B} - \vec{\theta}_{\rm A} \right)^{\rm T} \boldsymbol{\Sigma}_{\beta} \left(\vec{\theta}_{\rm B} - \vec{\theta}_{\rm A} \right).
\end{equation}

The propagation of a positional error covariance matrix in the image plane of image $k$, $\boldsymbol{\Sigma}_{\theta, k} \equiv \sigma(\theta_{k,i}, \theta_{k,j})$, to the source plane covariance, $\boldsymbol{\Sigma}_{\beta, k}$, is given by \citep[see e.g.][for the use of this relation for lens modelling]{Oguri:2010}

\begin{equation}\label{eqn:source_individual}
\boldsymbol{\Sigma}_{\beta, k} = \boldsymbol{A}_k^{\rm T} \boldsymbol{\Sigma}_{\theta, k} \boldsymbol{A}_k,
\end{equation}
where
\begin{equation}
\boldsymbol{A}_k \equiv \frac{\partial \vec{\beta}}{\partial \vec{\theta}_k}
\end{equation}
is the lensing Jacobian at the image position $k$. In the strong gravitational lensing regime, the  source plane is magnified and thus $\text{det}\left(\boldsymbol{\Sigma}_{\beta, k}\right) \le \text{det}\left(\boldsymbol{\Sigma}_{\theta, k} \right)$.

In case of uncorrelated astrometric errors of multiple images of a source, the Gaussian error propagation on the source uncertainty is given by

\begin{equation} \label{eqn:source_combined}
\boldsymbol{\Sigma}_{\beta} = \left( \sum_k \boldsymbol{\Sigma}^{-1}_{\beta, k} \right)^{-1},
\end{equation}
where $\boldsymbol{\Sigma}^{-1}_{\beta, k}$ are the inverse covariance matrices based on the individual images (Eqn \ref{eqn:source_individual}) summed over the multiple images (indexed by $k$). 

\begin{figure*}
  \centering
  \includegraphics[angle=0, width=180mm]{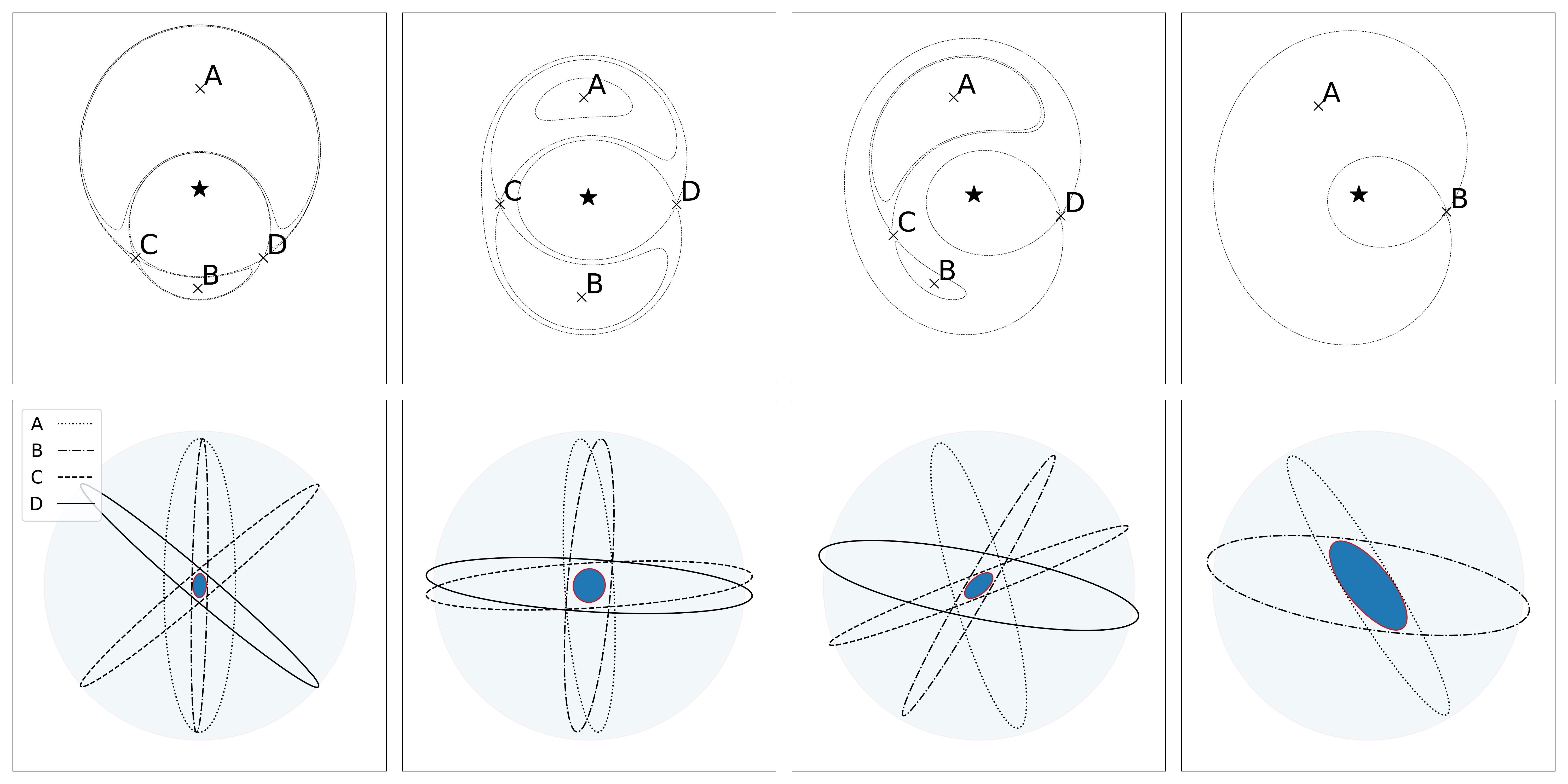}
  \caption{Illustration of the astrometric error propagation from the image to the source plane for three common quadruply imaged lensing configurations (cusp, cross, and fold configuration) and a doubly imaged lensing configuration. \textbf{Top panel:} Image positions and arrival time surfaces for the four source configurations. \textbf{Lower panel:} Error ellipses in the source plane from the individual images (Eqn \ref{eqn:source_individual}) in different line styles as indicated by the caption and combined (Eqn \ref{eqn:source_combined}) as solid blue contour with red edge. The shaded region corresponds to the astrometric precision in the image plane for an individual image. Different line styles correspond to the different individual images.}
  \label{fig:lens_examples}
\end{figure*}

Figure~\ref{fig:lens_examples} illustrates the astrometric error propagation from the image to the source plane for three common quadruply imaged lensing configurations (cusp, cross, and fold configuration) and a doubly imaged lensing configuration (upper panel of Figure \ref{fig:lens_examples}) for a singular isothermal ellipsoid with external shear lens model. The lower panel corresponds to the error ellipses in the source plane from the individual images (Eqn \ref{eqn:source_individual}) and combined (Eqn \ref{eqn:source_combined}). Quadrulpy imaged lensing configuration naturally provide more information about the source position than doubly imaged ones.

In Table~\ref{table:linear_example} we present numerical values of the error propagation of astrometric errors on the relative Fermat potential under fixed lensing potential (Eqn \ref{eqn:linear_fermat_errors}) of the four lensing configurations presented in Figure~\ref{fig:lens_examples}, assuming an Einstein radius $\theta_{\rm E} = 1\arcsec$ and a fiducial astrometric uncertainty of $\sigma_\theta = 10$ mas. The errors are given relative to the true relative Fermat potential. The last column presents the relative error propagation from the image to the source plane. The uncertainty in the source position is about one order of magnitude smaller than the astrometric uncertainty on individual image positions. We note that some relative Fermat potentials are affected at the $\sim$10 per cent level, even in the case of perfect knowledge of the lensing model.

\begin{table*}
\caption{Error propagation of astrometric errors on the relative Fermat potential under fixed lensing potential (Eqn \ref{eqn:linear_fermat_errors}) of the four lensing systems presented in Figure \ref{fig:lens_examples} with an astrometric error of 1 per cent relative to the Einstein radius $\theta_{\rm E}$ (e.g. 10 mas for $\theta_{\rm E} = 1 \arcsec$). The errors are given relative to the true relative Fermat potential. The last column presents the relative error propagation from the image to the source plane.}
\begin{center}
\begin{threeparttable}
\begin{tabular}{l r r r r r r r}
    
\hline
configuration & $\sigma_{\Delta \phi_{\rm AB}} / \Delta \phi_{\rm AB}$ & $\sigma_{\Delta \phi_{\rm AC}} / \Delta \phi_{\rm AC}$ & $\sigma_{\Delta \phi_{\rm AD}} /\Delta \phi_{\rm AD}$ & $\sigma_{\Delta \phi_{\rm BC}} / \Delta \phi_{\rm BC}$ & $\sigma_{\Delta \phi_{\rm BD}} / \Delta \phi_{\rm BD}$ & $\sigma_{\Delta \phi_{\rm CD}} / \Delta \phi_{\rm CD}$ & $\sigma_\beta / \sigma_\theta $  \\ 
\hline
cusp & 0.01 & 0.01 & 0.01 & 0.06 & 0.05 & 0.68 & 0.06  \\ 
cross & 0.11 & 0.02 & 0.02 & 0.02 & 0.02 & 0.1 & 0.11  \\ 
fold & 0.02 & 0.02 & 0.01 & 0.06 & 0.02 & 0.02 & 0.08  \\ 
double & 0.03 & - & - & - & - & - & 0.22  \\ 
\hline
\end{tabular}
\begin{tablenotes}
\end{tablenotes}
\end{threeparttable}
\end{center}
\label{table:linear_example}
\end{table*}

\subsection{Non-linear error propagation}\label{sec:astrometry_non_linear}
The calculation in Section~\ref{sec:astrometry_first_order} assumes that the astrometric uncertainty does not affect the precision of the inference of the lensing potential $\psi$. This is true in the regime where extended arc and ring features dominate the information used to constrain the lens model. In this section, we perform an error propagation in the case where the image positions are the dominant source of information of the lens model itself. As an illustration, we choose the same three lensing configurations presented in Figure~\ref{fig:lens_examples} and Section~\ref{sec:astrometry_first_order}.

We perform the following Monte Carlo error propagation for all four cases:
\begin{enumerate}
  \item We draw a displacement from a Gaussian error distribution with $\sigma_\theta$ for each of the 4 (2) images.
  \item We re-adjust the lens model parameters to satisfy the constraints on the image positions of our imperfect measurement. To find an unique solution for positions, we fix the shear strength for the quads to the true value and for the double, providing far fewer constraints on the lens model, we fix the lensing center and orientation of the deflector ellipticity as well (possibly informed by a luminous deflector component).
  \item We compute the relative Fermat potential at the observed image positions with the re-adjusted lens model.
  \item We repeat these steps to produce a posterior distribution of the relative Fermat potential.
\end{enumerate}
As a result, we have a posterior distribution of relative Fermat potentials representing the uncertainty of the astrometry. Table~\ref{table:non_linear_example} presents the errors on the relative Fermat potential for the four cases.
The source position uncertainty is comparable to the image plane astrometric uncertainty of the individual images, significantly larger relative to the linear error propagation when assuming the correct lens model (Table \ref{table:linear_example}). The relative Fermat potential uncertainties are inflated by the same order as the source position uncertainties.
The specific error on the relative Fermat potential depends on the image configuration and possibly also on the lens model assumptions. Even though the details may vary for each system, the scaling of the uncertainties with source position uncertainty and image separation according to Equation \ref{eqn:linear_fermat_errors} should be robust in most practical cases. We do not observe a bias in the relative Fermat potentials when incorporating random Gaussian errors in the astrometry.

\begin{table*}
\caption{Error propagation of astrometric errors on the relative Fermat potential of the three lensing systems presented in Figure \ref{fig:lens_examples} while simultaneously inferring the lensing potential from the image positions with an astrometric error of 1\% relative to the Einstein radius $\theta_{\rm E}$ (e.g. 10 mas for $\theta_{\rm E} = 1 \arcsec$). The errors are presented relative to the true relative Fermat potential. The last column presents the relative error propagation from the image to the source plane.}
\begin{center}
\begin{threeparttable}
\begin{tabular}{l r r r r r r r}
    
\hline
configuration & $\sigma_{\Delta \phi_{\rm AB}} / \Delta \phi_{\rm AB}$ & $\sigma_{\Delta \phi_{\rm AC}} / \Delta \phi_{\rm AC}$ & $\sigma_{\Delta \phi_{\rm AD}} /\Delta \phi_{\rm AD}$ & $\sigma_{\Delta \phi_{\rm BC}} / \Delta \phi_{\rm BC}$ & $\sigma_{\Delta \phi_{\rm BD}} / \Delta \phi_{\rm BD}$ & $\sigma_{\Delta \phi_{\rm CD}} / \Delta \phi_{\rm CD}$ & $\sigma_\beta / \sigma_\theta $  \\ 
\hline
cusp & 0.26 & 0.27 & 0.27 & 0.32 & 0.32 & 0.96 & 0.63  \\ 
cross & 0.18 & 0.13 & 0.13 & 0.13 & 0.14 & 0.23 & 0.72  \\ 
fold & 0.18 & 0.18 & 0.2 & 0.21 & 0.23 & 0.23 & 0.71  \\ 
double & 0.18 & - & - & - & - & - & 0.74  \\
\hline
\end{tabular}
\begin{tablenotes}
\end{tablenotes}
\end{threeparttable}
\end{center}
\label{table:non_linear_example}
\end{table*}

\subsection{Astrometric requirements for cosmography} \label{sec:astrometry_requirement}
In Section~\ref{sec:astrometry_first_order}, we derived a first order error propagation of astrometric uncertainties on the image positions of variable sources and concluded that the source position uncertainty is the dominant first order term when the lens model is perfectly known (or known from other constraints like extended images).
In contrast, when using the image positions as constraints on the lens model itself (Section~\ref{sec:astrometry_non_linear}), we concluded that the uncertainty in the source position is of order the image position uncertainties. This statement applies to all three configurations (cusp, cross, and fold) with the same linear scaling of the astrometric precision applicable as derived from the linear error correction terms.

In this section, we use the astrometric error formula (Eqn \ref{eqn:error_vector}, \ref{eqn:linear_fermat_errors}) to derive astrometric requirements based on the observed image separations and relative time delays for current and future lens systems.

We can express the error propagation of Equation~\ref{eqn:error_vector} in terms of measured time delays, $\Delta t_{\rm AB}$, on the Hubble constant (Eqn \ref{eqn:hubble_fermat_error}) as

\begin{equation}
\frac{\delta {\rm H}_0(\delta \vec{\theta})}{{\rm H}_0} \approx
\frac{D_{\Delta t}}{c \Delta t_{\rm AB}} \left(\vec{\theta}_{\rm B} - \vec{\theta}_{\rm A} \right) \cdot \delta \vec{\beta}.
\end{equation}

For deriving astrometric requirements, we further simplify the equation above by removing the vector equation resulting in the expression
\begin{equation} \label{eqn:error_h0}
\frac{\sigma_{{\rm H}_0}}{{\rm H}_0} \approx
\frac{D_{\Delta t}}{c} \frac{\theta_{\rm AB}}{\Delta t_{\rm AB}} \sigma_\beta.
\end{equation}

In order for the astrometric uncertainty to be sub-dominant with respect to the uncertainty in the time delay measurement, $\sigma_{\Delta t_{\rm AB}}$, the following requirement applies (from Eqn \ref{eqn:hubble_dt_error})

\begin{equation} \label{eqn:requirement_measurement}
\theta_{\rm AB} \sigma_\beta \lesssim \sigma_{\Delta t_{\rm AB}} \frac{c}{D_{\Delta t}}.
\end{equation}
For a fixed time-delay precision, the requirements are more stringent for larger separation lenses.
However, the time-delays scale with lens size as $\Delta t_{\rm AB} \propto \theta_{\rm AB}^2$ and the error on the Hubble constant thus scales as
\begin{equation}
  \sigma_{{\rm H}_0} \propto \frac{\sigma_\beta}{\theta_{\rm AB}}.
\end{equation}
In other words, the astrometric requirements for small separation short time-delay lenses are more stringent than for the same symmetry with larger image separations to achieve the same precision on the Hubble constant. This is on top of the fact that small separation lenses already have shorter time delays, and hence a larger time delay uncertainty propagated into the Hubble constant (e.g. Eqn \ref{eqn:hubble_dt_error}).
The values for the relative Fermat potential difference errors in Table~\ref{table:linear_example} and \ref{table:non_linear_example} can be re-scaled by the factor $\sigma_\theta / \theta_{\rm E}$ for the full dynamic range of gravitational lensing.
To relate the requirements from the source plane, $\sigma_\beta$, to the image plane astrometric uncertainty, $\sigma_\theta$, the number of images and their individual magnifications have to be taken into account (Eqn \ref{eqn:source_combined}). This leads to $\sigma_\theta / \sigma_\beta \sim 10^{-1}$ (Table \ref{table:linear_example} last row). In case when the positional information is used to determine the lens model itself, the non-linear error propagation results in $\sigma_\theta / \sigma_\beta \sim 1$ (Table \ref{table:non_linear_example} last row).

\section{Specific astrometric requirements} \label{sec:specific_requirements}

In the following section, we illustrate the astrometric precision requirements given by Equations~\ref{eqn:error_h0} and \ref{eqn:requirement_measurement} with some specific examples of image separations, measured time delay and its precision, that are meant to describe typical conditions for present-day and near-future campaigns. When needed, we assume a flat $\Lambda$CDM cosmology with H$_0 = 70$ km s$^{-1}$ Mpc $^{-1}$ and $\Omega_{\rm m} = 0.3$. We set the deflector redshift $z_{\rm d} = 0.5$ and source redshift $z_{\rm s} = 2$. We note that different and equally reasonable choices of redshift configuration and background cosmology would only marginally modify the requirements.
We present specific examples in Section \ref{sec:examples} and discuss the major challenges identified by our work in Section \ref{sec:challenges}.
These requirements have to be matched by the precision a lens model predicts the observed image positions.

\begin{table*}
\caption{Astrometric requirements for five different examples of image separations, $\theta_{\rm AB}$, and time delays, $\Delta t_{\rm AB}$, at lens redshift z$_{\rm d} = 0.5$ and source redshift $z_{\rm s} = 2$. The requirements are for uncertainty on H$_0$ from astrometric uncertainty to be less than the uncertainty on H$_0$ from the time delay uncertainty, $\sigma_{\Delta t}$, and to be below the 5 per cent level, which we choose as representative of the total uncertainties including those arising from modelling the main deflector potential and the contribution from the mass along the line of sight. The requirements are expressed as uncertainties in the source position, $\sigma_\beta$. To relate these requirements to the image plane astrometric uncertainty, $\sigma_\theta$, the number of images and their individual magnifications have to be taken into account (Eqn \ref{eqn:source_combined}). This leads to $\sigma_\theta / \sigma_\beta \sim 10^{-1}$ (Table \ref{table:linear_example} last row). In case when the positional information is used to determine the lens model itself, the non-linear error propagation results in $\sigma_\theta / \sigma_\beta \sim 1$ (Table \ref{table:non_linear_example} last row).}
\begin{center}
\begin{threeparttable}
\begin{tabular}{l r r r r r}

    \hline
    Example & $\theta_{\rm AB}$ [arcsec] & $\Delta t_{\rm AB}$ [days] & $\sigma_{\Delta t}$ [days]&  $\sigma_{{\rm H}_0}(\sigma_\beta) \le \sigma_{{\rm H}_0}(\sigma_{\Delta t}$) [mas] &  $\sigma_{{\rm H}_0}(\sigma_\beta) \le 5\%$ [mas]  \\
    \hline
1 & 20 & 1000 & 30 & 18 & 30  \\ 
2 & 3 & 100 & 3 & 12 & 20  \\ 
3 & 2 & 10 & 1 & 6 & 3  \\ 
4 & 1 & 4 & 0.25 & 3 & 2.4  \\ 
5 & 1 & 1 & 0.025 & 0.3 & 0.6  \\ 
\hline
\end{tabular}
\begin{tablenotes}
\end{tablenotes}
\end{threeparttable}
\end{center}
\label{table:requirements_list}
\end{table*}

\subsection{Examples} \label{sec:examples}
Table~\ref{table:requirements_list} gives five specific examples of image separation and time delays and their required astrometric precision to (a) match the quoted time-delay precision (Eqn \ref{eqn:requirement_measurement}) and (b) to match a 5 per cent uncertainty in H$_0$ not to dominate the error budget of the cosmographic analysis in respect to other errors (Eqn \ref{eqn:error_h0}). The time-delay measurement precisions are chosen to match current observational precision (\textit{Example 1-3}) or future ambitious measurements (\textit{Example 4 \& 5}).
In the following, we discuss those examples.

\textit{Example 1:}
For a typical cluster-scale lens with image separation of 20\arcsec and a time delay of 1000 days, the relative astrometric requirement is 30 mas in the source plane to not exceed a 5 per cent uncertainty in H$_0$. This is routinely achieved with current \textit{HST} imaging. A distortion inaccuracy of 6 mas over the scale of the cluster \citep[][]{Kozhurina-Platais:2018} results in a 1 per cent effect on H$_0$. Unaccounted distortions do have an impact on the lens model and thus are non-linear effects (Section \ref{sec:astrometry_non_linear}). In the case of supernova "Refsdal" in MACSJ1149+2223 ($z_{\rm d} = 0.54$ and $z_{\rm s} = 1.49$) the image separation of SX and S1-4 is about 8\arcsec and the time delay is about 350 days \citep{Kelly:2016}. Thus the astrometric requirement in the source plane is approximately 20 mas to match a 5\% precison on H$_0$ from a time delay of SX. The magnification of SX is estimated to be about 5 \citep{Grillo:2016} leading to a required image plane precision of about 50 mas (with directional dependence). The fully non-linear propagated uncertainties of the \cite{Grillo:2016} models achieving of order 100 mas r.m.s precision in the position of SX and sub 10\% precision \citep{Grillo:2018} in the relative time delay confirm the validity of the estimates provided by the framework presented in this work.

\textit{Example 2:}
A typical massive galaxy with image separations of 3\arcsec and a relative time delay of 100 days requires a similar relative astrometric precision of 20 mas in the source plane to match the 5 per cent uncertainty requirement on H$_0$. This is similar to the case of the longest time delays in the quadruply lensed quasars RXJ1131-1231 \citep{Suyu:2013ni, Suyu:2014aq, Birrer:2016zy} and B1608+656 \citep{Suyu:2010rc}, or for the doubly lensed quasar SDSSJ1206+4332 \citep{Birrer:2019_1206}.

\textit{Example 3:}
Smaller separation images of 2\arcsec  with a relative time delay of 10 days lead to a 3 mas astrometric requirement, significantly more demanding than examples 1 and 2.
This example corresponds to either neighboring images of a wide separation quadruply lensed quasars or a smaller separation lens. This case is similar to HE 0435-1223 \citep{Wong:2017} or the smaller time-delay pairs in RXJ1131-1231 and B1608+656.

\textit{Example 4:}
Short time delays with image separation of 1\arcsec and a relative delay of 4 days, as expected from small separation lenses or neighboring image pairs must meet an astrometric requirement of 2.4 mas.

\textit{Example 5:}
Small and symmetric lenses with image separation of 1\arcsec and a relative delay of 1 day results in sub mas requirements in the relative astrometry to reach a 5\% precision measurement. Realistic time-delay measurements will be less precise than quoted in Table \ref{table:requirements_list} unless other measurements (such as gravitational waves) are available. This last example is motivated by the recent discovery of the lensed supernova iPTF16geu \citep{Goobar:2017} and X-ray measurements of the close pair in PG1115+080 \citep{Chartas:2004}. An astrometric uncertainty in the source plane of 12 mas would result in an uncertainty of the Hubble constant of unity.

\subsection{Challenges} \label{sec:challenges}

For galaxy clusters (e.g. \textit{Example 1} in Table \ref{table:requirements_list}) the astrometric precision is not a limiting factor. However, the complexity in the lens model and their uncertainties may limit the predictive power of where the images appear and thus are limited in the ability to infer the source position. 
The lens models are mostly informed by conjugate points other than the actual variable source images and thus the linear propagation of uncertainties from the image to the source plane is applicable.
A limiting factor in a cluster analysis is the ability of the lens model to map the source to the correct image positions. In this regard, a source position uncertainty of 30 mas is required to measure H$_0$ to the 5 per cent level. This translates to an image position uncertainty of e.g. 60 mas by an image magnified by a factor of four. Achieving this precision is challenging in the cluster lensing regime and requires a dedicated effort and exquisite data \citep{Grillo:2018}, beyond what is usually done in cluster models, where root-mean-square (r.m.s.) scatter of positions in the image plane are typically at the $0.1-1\arcsec$ level \citep{Treu:2016Refsdal}.

The astrometric requirements of few mas are at the limit of the \textit{HST} capabilities for isolated point sources. In the strong lensing systems considered here, the variable images are in a crowded field, superposed on a host galaxy and stellar light from the lensing galaxy. This makes precise astrometry more challenging than for isolated point sources. Achieving astrometric precision of 10 mas with HST requires a considerable and deliberate effort in the reconstruction of the point spread function, the design of the observations for optimal sampling of the data and the accuracy in the modeling of different blended light components.
Centroiding and interpolating point spread functions in the forward modeling of imaging data and we advise that those limitations are quantified and formally propagated through the lens modeling analysis in regimes where impacts on the results are expected from the stated requirements.
For radio loud sources, radio interferometry may turn out to be very useful in the most extreme cases. The main caveat is that combining radio positions with time delays from optical monitoring leads to biased results if the radio and optical emission are not coincident \citep[see e.g.][for a recent review of the potential misalignment of the nuclear emission at different wavelengths]{Barnacka:2018}. \textit{GAIA} astrometry has also its limitation due to the blending of galaxy host and image components \citep[see e.g.][]{Krone-Martins:2018}. Large ground-based telescopes equipped with adaptive optics like Keck and the proposed next generation of extremely large telescopes like the Thirty Meter Telescope \citep{Sanders:2013}, Giant Magellan Telescope \citep{Bernstein:2014} or the European Extremely Large Telescope will be able to meet these requirements taking advantage of their resolution and well-sampled PSFs, provided that the PSF and its variations can be faithfully reconstructed over the field of view of the lens.

Additionally, dark matter substructure is known to perturb the image positions by several mas \citep{Chen:2007}. These lensing perturbations result in a fundamental limit on how precisely a specific lens model can predict the image positions and that requirements of few mas or below may be unattainable given the unresolved lensing perturbation at those scales.

Finally, we note that astrometric uncertainties introduce an additional noise term in the measurement of time delay anomalies \citep{Keeton:2009}. Although investigating that effect is beyond the scope of this work, the formulae given here can be used for that purpose as well.

\section{Summary} \label{sec:conclusion}

We investigated the effect on time-delay cosmography of astrometric errors on the point-like images of time variable sources such as quasars or supernovae. We derived a convenient analytic expression (Eqn \ref{eqn:error_h0}) for the propagation of astrometric uncertainties on the multiple image positions into the inference of the Hubble constant depending on image separation and relative time delay. We derived requirements for different type of lenses so that the astrometric uncertainty is sub-dominant with respect to other uncertainties in the cosmographic analysis.

Meeting the derived astrometric requirements is essential when performing a cosmographic analysis. For the current wide-separation galaxy-scale quasar lenses with relative time delays of about 100 days, the requirements are met with \textit{HST} imaging. For transients that are more challenging to localize (such as gravitational waves), astrometric uncertainties can be a major limitation.

In the cluster regime, whereas the nominal astrometric precision can easily be achieved by \textit{HST} images, the requirement of the lens model to match multiple images with mas precision translates into
a source position uncertainty which poses a strong limitation on the cosmographic analysis. The same formulae derived in this paper for astrometric uncertainties can be used to assess the uncertainty in cosmography stemming from the r.m.s. residuals between the observed and predicted image positions from a given lens model.

For small time-delay lenses with relative delays in the regime of $1-10$ days, the astrometric requirements tighten to order mas. Meeting these requirements is a major challenge due to crowding by the host and lens galaxy with current imaging capacity and may require radio interferometry or adaptive optics behind the next generation of extremely large telescopes in the near infrared.

Ultimately, dark matter substructure introduces astrometric anomalies at the mas level (known as millilensing), that may set a fundamental noise floor per system, particularly relevant for small-separation short-delay lenses. It is beyond the scope of this paper to investigate whether this noise term can be contained by averaging over many systems. 

Taking the broader view, we recommend that the efforts to increase the precision of time-delay measurements and lens modelling techniques must be matched with corresponding efforts to increase astrometric precision and the modelling aspects of it. For example, in the setting of the Time-Delay Lens Modelling Challenge \citep{Ding_TDLMC}, an optimal measurement of the time delays of 0.25 days is adopted, corresponding to the ideal case of a high-cadence high-precision monitoring campaign. In such a case, astrometry may dominate the error budget and become a limitation, especially for small-separation short time-delay lenses.

\section*{Acknowledgments}

SB thanks Martin Millon and Aymeric Galan to help identifying the issue with astrometry while working on the Time-delay Lens Modelling Challenge \citep{Ding_TDLMC}. We thank Adriano Agnello, Anowar Shajib, Chih-Fan Chen, Dominique Sluse, Patrick Kelly, Xuheng Ding and Jenny Wagner for useful feedback on the manuscript. We thank the anonymous referee for useful suggestions and comments that improved the presentation of this work. Support for this work was provided by NASA through grant numbers HST-G0-14254 and HST-G0-15320 from the Space Telescope Science Institute, which is operated by AURA, Inc., under NASA contract NAS 5-26555. TT acknowledges support by the NSF through grant AST-1714953 and by the Packard Foundation through a Packard Research Fellowship.




\bibliographystyle{mnras}
\bibliography{BibdeskLib}{}



\appendix

\bsp    
\label{lastpage}
\end{document}